\documentclass[12pt,preprint]{aastex}

\def\oiii{\ifmmode [O {\sc iii}] \else [O {\sc iii}]\ \fi}
\def\oii{\ifmmode [O {\sc ii}] \else [O {\sc ii}]\ \fi}
\def\Nii{\ifmmode [O {\sc ii}] \else [N {\sc ii}]\ \fi}
\def\Neii{\ifmmode [O {\sc ii}] \else [Ne {\sc ii}]\ \fi}
\def\Neiii{\ifmmode [O {\sc iii}] \else [Ne {\sc iii}]\ \fi}
\def\Nev{\ifmmode [O {\sc v}] \else [Ne {\sc v}]\ \fi}
\def\oiv{\ifmmode [O {\sc iv}] \else [O {\sc iv}]\ \fi}
\def\oi{\ifmmode [O {\sc i}] \else [O {\sc i}]\ \fi}
\def\sii{\ifmmode [O {\sc ii}] \else [S {\sc ii}]\ \fi}
\def\Hii{\ifmmode [O {\sc ii}] \else H{\sc ii}\ \fi}

\def\aj{\rm{AJ}}                   
\def\araa{\rm{ARA\&A}}             
\def\apj{\rm {ApJ}}                
\def\apjs{\rm{ApJS}}               
\def\aap{\rm{A\&A}}                
\def\aaps{\rm{A\&AS}}              
\def\mnras{\rm{MNRAS}}             

\usepackage{graphicx}
\begin{document}
\title{Evolution of oxygen and nitrogen abundances and nitrogen production
mechanism in massive star-forming galaxies}
\author{Yu-Zhong Wu\altaffilmark{1} and Shuang-Nan Zhang\altaffilmark{1,}\altaffilmark{2}}
\altaffiltext{1}{Key Laboratory for Particle Astrophysics, Institute of High Energy Physics, Chinese Academy of
Sciences, 19B Yuquan Road, Beijing 100049 (zhangsn@ihep.ac.cn)} \altaffiltext{2}{National Astronomical
Observatories, Chinese Academy of Sciences, 20A Datun Road, Beijing 100012} \shorttitle{Abundance in massive star formation galaxies} \shortauthors{Wu $\&$ Zhang} \slugcomment{}

\begin{abstract}

Utilizing the observational data of 55,318
star-forming galaxies (SFGs) selected from the catalog of MPA-JHU
emission-line measurements for the SDSS DR8, we investigate the galaxy downsizing effect of their O and N
enrichments, and the nitrogen production mechanism in them. We show
the redshift evolution of O and N abundances and specific star formation rates for different galaxy
mass ranges, demonstrating
the galaxy downsizing effect caused by less massive progenitors of less massive galaxies. The O and N
abundances do not remain constant for different galaxy mass ranges,
and the enrichment (and
hence star formation) decreases with increasing galaxy
stellar mass. We find evidence of the O enrichment for galaxies with
stellar masses $M_{*}>10^{11.0} $ (in units of $M_{\odot}$), i.e.
$\Delta({\log}({\rm O/H})) \sim 0.10$ and $\Delta({\log}({\rm N/H}))
\sim 0.28$ from redshift 0.023 to 0.30. Based on the evolutionary
schematic model of N/O ratios in Coziol et al., who proposed the scheme that the production of nitrogen is
the consequence of a sequence of bursts in SFGs, we conclude that
the nitrogen production is dominated by the intermediate-mass stars,
which dominate the secondary synthesis in SFGs. However, for
galaxies with $M_{*}>10^{10.35} $ we find evidence of enhanced N/O
abundance ratios, which are significantly above the secondary
synthesis line. This suggests that outflows of massive stars, which deplete oxygen efficiently, are more important in massive galaxies. Finally we find an
excellent linear relation between $M_{*}$ and log(N/O), indicating
that the N/O abundance ratio is a good indicator of the stellar mass
in a SFG and may be used as a standard candle for studying
cosmology, if confirmed with further studies.

\end{abstract}

\section{INTRODUCTION}

The metallicity of a galaxy is a crucial parameter for understanding
its formation and evolution. The element abundances of the
interstellar medium (ISM) are obtained by tracers of the chemical
compositions of stars and gas within a galaxy. Optical emission
lines from \Hii regions have long been regarded as the principal
tools of gas-phase chemical diagnostics in galaxies (Aller 1942;
Peimbert 1975; Pagel 1986; de Robertis 1987; Liang et al. 2006).
Since estimating metallicities needs theoretical models, empirical
calibrations or a combination of both (Kewley \& Dopita 2002; Kewley
\& Ellison 2008), we have different methods to obtain metallicities.

Assuming a classical \Hii region model, the ratio of the auroral
line \oiii $\lambda4363$ to a lower excitation line such as
\oiii$\lambda 5007$, can be used to determine the electron
temperature of the gas, which is then converted into the metallicity
of the gas (Osterbrock 1989). The method of using the observed line
ratios to infer directly the electron temperatures and to estimate
metallicities in galaxies is known as ``direct $T_{\rm e}$ method"
(Pagel et al. 1992; Skillman \& Kennicutt 1993). This method is
generally regarded as the most accurate abundance measurement for
estimating metallicities in galaxies. It, however, has two
disadvantages. In most instances, \oiii $\lambda 4363$ line is too
weak to be detected. It is well known that in metal-rich galaxies,
the electron temperature decreases (as the cooling is via metal
lines) and the auroral lines eventually become too faint to be
measured, when the metallicity increases (Yin et al. 2007). In
addition, in low-metallicity galaxies, the oxygen abundance is
usually underestimated by the \oiii $\lambda 4363$ diagnostic
in low-metallicity galaxies (Kobulnicky et al.
1999).

Due to the above reasons, photoionization models are used instead
for estimating abundances of high metallicity star-forming galaxies
(SFGs). The most wide and common method used is the R23 method
proposed by Pagel et al. (1979) and Alloin et al. (1979); the oxygen
indicator $R_{23}$=(\oii $\lambda \lambda 3727, 3729 +$ \oiii
$\lambda \lambda 4959, 5007)/\rm H\beta$ suggested by Pagel et al.
(1979) is widely accepted and used. Moreover, the relation between
the line intensities of strong oxygen lines and the oxygen abundance
has been calibrated by photoionization models (e.g., Edmunds \&
Pagel 1984; McCall et al. 1985; Dopita \& Evans 1986; Kobulnicky et
al. 1999; Kewley \& Dopita 2002). However, it has one problem that
the relationship between $R_{23}$ and $12+{\log} ({\rm O/H})$ is
double valued, which shows the transition between the upper
metal-rich branch and the lower metal-poor branch occurring near
$12+\rm log (O/H)\sim8.4$ (Liang et al. 2006).

With the releases of catalogues of several large spectral surveys,
especially the Sloan Digital Sky Survey (SDSS) that has released a
large number of spectral data, the number of good-quality spectra of
emission-line galaxies has increased dramatically (York et al.
2000). These open a new era of utilizing the large survey spectra to
study the evolution of O and N abundances in galaxies. Using the
line flux measurements of SDSS spectra, Thuan et al. (2010) not only
found the evolution of O and N abundances in galaxies with different
stellar masses, but also found evidence for galaxy downsizing that
metal enrichment shifts from higher-mass galaxies at early cosmic
times to lower-mass ones at later epochs (Cowie et al. 1996).

In the last decade, the evolution of the mass-metallicity relation
of galaxies with redshift has been investigated by several groups
(Lilly et al. 2003; Savaglio et al. 2005; Erb et al. 2006; Cowie \&
Barger 2008; Maiolino et al. 2008; Lamareille et al. 2009;
Lara-L\'{o}pez et al. 2009). In these studies, they used different
methods to obtain O abundances and found the O abundance change of
SFGs, with $\Delta ({\log} ({\rm O/H}))\sim0.3$ or lower. In
addition, Thuan et al. (2010) have paid attention to the redshift
evolution of N abundances in galaxies and have shown two advantages
for studying the chemical evolution of galaxies. Firstly, at
$12+({\log} ({\rm O/H}))\gtrsim8.3$, nitrogen abundance change with
redshift has a larger amplitude than that of oxygen. Then, compared
with oxygen production, the nitrogen production has a time delay
(Maeder 1992; van den Hoek \& Groenewegen 1997; Pagel 1997), and it
can give some limits for the chemical evolution of galaxies. Using a
large sample of galaxies in the Great Observatories Origins Deep
Survey-North (GOODS-N), Cowie \& Barger (2008) found that star
formation ceases in most massive galaxies (with stellar masses
$M_*>10^{11}$ in units of $M_{\sun}$) at $z<1.5$. Thuan et al.
(2010) also found $\Delta ({\log} ({\rm O/H}))=0$ for massive
galaxies with $M_*>10^{11}$.

\begin{figure}
\begin{center}
\includegraphics[width=10cm]{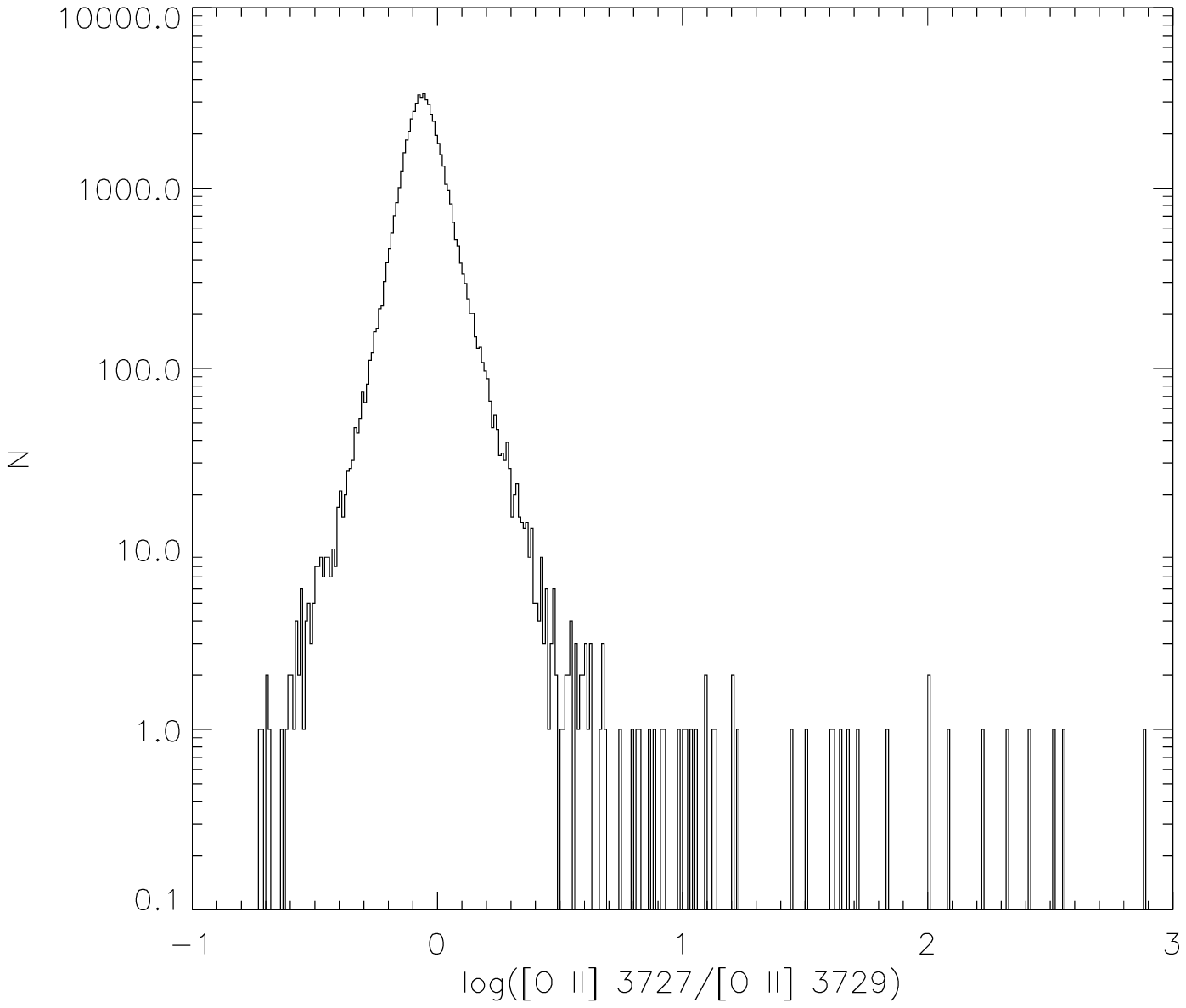}
\caption{The distribution of \oii$\lambda3727$/\oii$\lambda3729$
flux ratios.}
\end{center}
\end{figure}

Although the evolution of O and N abundances of galaxies with
redshift has been widely studied, we still cannot fully understand
the downsizing effect (i.e., some disputes for the origin of the
downsizing; Poggianti et al. 2004; Bundy et al. 2006). Moreover,
some studies seem to show no signs of the evolution of O abundance
with redshift in most massive galaxies ($M_*>10^{11}$). Therefore,
we utilize the MPA-JHU DR8 release of spectral
measurements to investigate these issues.

In this work, we first present the galaxy downsizing effect; we then
show evidence of the O enrichment for galaxies with $M_{*}>10^{11.0}
$. On the basis of the evolutionary schematic model of N/O ratios,
we can clarify the nitrogen production mechanisms in SFGs.

\begin{figure}
\begin{center}
\includegraphics[width=10cm]{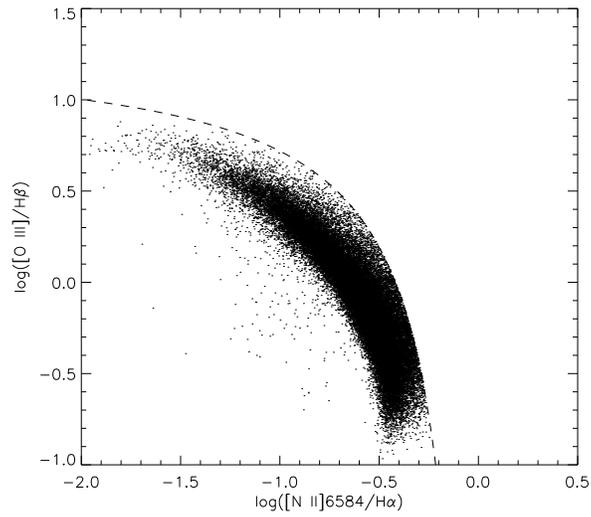}
\caption{Traditional diagnostic diagram. The dashed curve on the
diagram is the Kauffmann et al. (2003a) semi-empirical lower boundary
for the star-forming galaxies.}
\end{center}
\end{figure}

For the line fluxes, we use the following notations throughout this paper: $R_{\rm 2} = I_{\rm [O~II]
\lambda3727+\lambda3729}/I_{\rm H\beta}$, $R_{3}=I_{\rm[O~III] \lambda 4959+\lambda5007}/I_{\rm H\beta}$,
$N_{2}=I_{\rm[N~II] \lambda 6548+\lambda6584}/I_{\rm H \beta}$, $R_{\rm 23}=R_{\rm 2}+R_{\rm 3}$.

\section{THE SAMPLE and DATA}

The catalog of MPA-JHU emission-line measurements for the SDSS Data
Release 8 (DR8) is chosen as our primary sample. The
measurements are available for 1,843,200 different spectra. Compared
to the previous DR4 release used for a similar study (Wu, Zhao, \& Meng
2011), this represents a significant extension in
size and a number of improvements in the data.

The MPA-JHU DR8 release of spectral measurements
provides many parameters, such as the redshift, galaxy stellar mass,
and various emission line measurements. These data are available
online at the following address:
http://www.sdss3.org/dr8/spectro/galspec.php. We initially
select 71,888 objects
 from the above sample with the
criterion of S/N $>$ 5 for the redshift, H$\beta~\lambda 4861$,
H$\alpha~\lambda 6563$, \oii $\lambda 3727$, \oii $\lambda 3729$,
\Nii $\lambda 6584$, \oiii $\lambda 5007$, \oi $\lambda 6300$, \sii
$\lambda 6717$, and \sii $\lambda 6731$ lines. In addition,
we have excluded these galaxies for their FLAG keywords of zero and
$-9999.0$ in SFR and sSFR.

Due to that $3800${\AA}$-9300${\AA} is the wavelength range of the
SDSS spectra, the range of redshifts in our nearby galaxies can be
determined. If the \oii $\lambda \lambda 3727, 3729$ emission lines
are not out of the observed range, then the lower limit of redshifts
in our nearby galaxies is $z\approx0.023$. As shown in Fig. 1, a
tiny fraction of the objects show anomalously large
\oii~$\lambda3727$/\oii~$\lambda3729$ ratios; we thus exclude 30 objects with \oii~$\lambda3727$/\oii~$\lambda3729>10$.
If the \sii $\lambda \lambda 6717, 6731$ emission lines are not out
of the observed range, then the upper limit of redshifts in our
nearby galaxies is $z\approx0.33$ (Pilyugin \& Thuan 2011).

The SFG sample presented in Fig. 2 is based on the BPT diagram
(Baldwin et al. 1981; Veilleux \& Osterbrock 1987), and the
separation line between AGNs and SFGs proposed by Kauffmann et al.
(2003a) is the criterion of the SFG sample. We select all the
objects which reside in the lower left corner of the Kauffmann et
al. (2003a) line as our SFG sample, and we exclude the objects with
redshift $z>0.30$. Therefore, we obtain the galaxy sample with
55,318 objects.

In addition to the redshift of each galaxy from the release of the
MPA-JHU DR8 catalog, the release also provides the stellar masses of all galaxies
based on the fits to the photometry following the philosophy of
Kauffmann et al. (2003a) and Salim et al. (2007); the stellar
masses are FITS binary tables with keys MEDIAN. The galaxy stellar
masses come all from the MPA-JHU DR8 catalog in this paper. In
addition, the specific star formation rate (sSFR), taken as the median of the PDF
of the sSFR for each SFG, is also added for
completeness.

\begin{figure}
\begin{center}
\includegraphics[width=10cm]{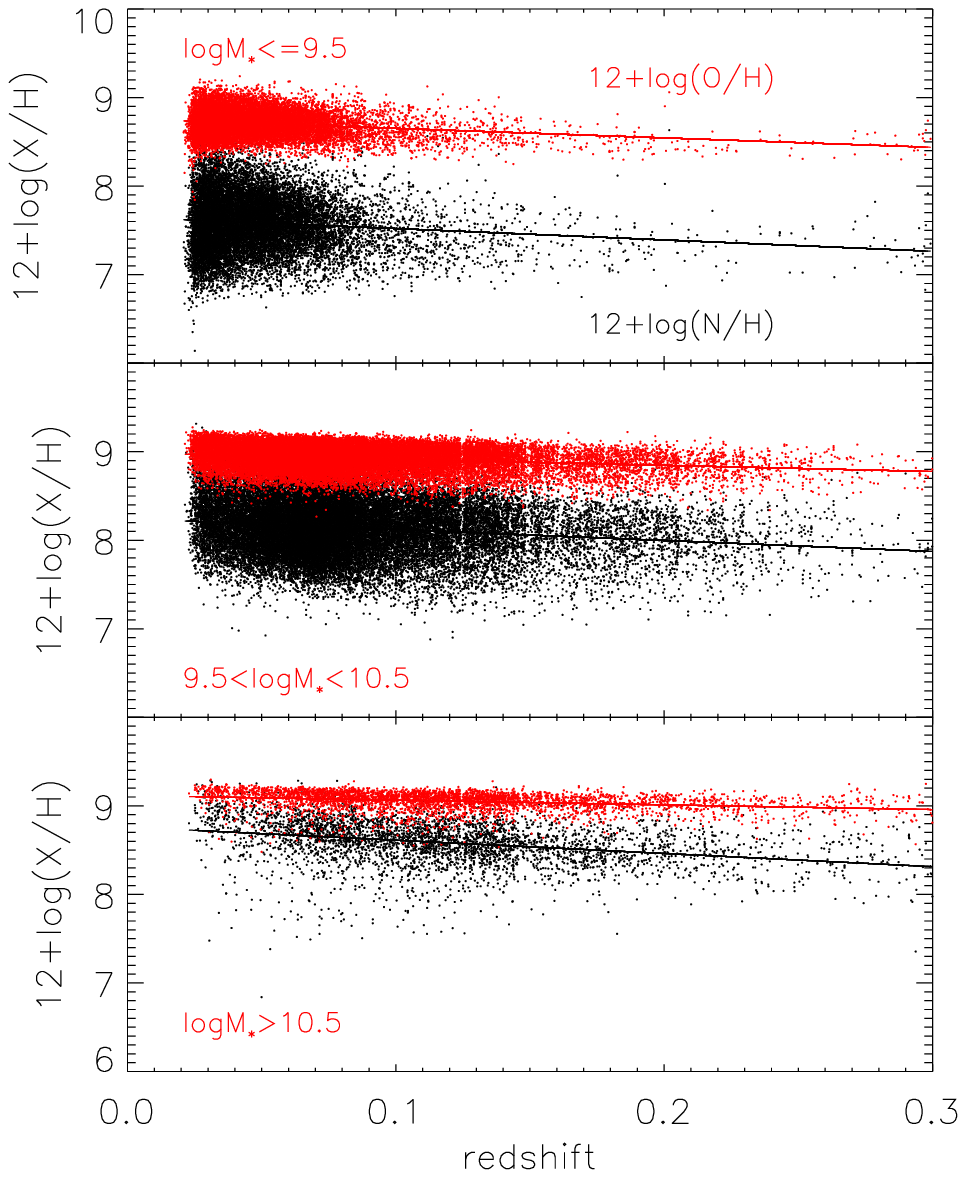}
\caption{Oxygen and nitrogen abundances as a function of redshift for those galaxies with $M_{*} < 10^{9.5} $,
$10^{9.5}  - 10^{10.5} $, and $ > 10^{10.5} $. Oxygen and nitrogen abundances are shown by red and black points,
respectively. The red and black solid lines are the best least-squares fits for these data.}
\end{center}
\end{figure}

\section{Metallicity calibrations}

In this section, we estimate the oxygen and nitrogen abundances
utilizing both emission-line fluxes and recent calibrations. The
oxygen abundances of the sample galaxies are estimated using the
$R_{23}$ method (Pilyugin et al. 2006; Pilyugin et al. 2010; Zahid
et al. 2012),
\begin{equation}
R_{\rm 23}= R_{\rm 2}+R_{\rm 3}.
\end{equation}

We adopt the calibration of Tremonti et al. (2004),
\begin{equation}
12+{{\log}({\rm O/H})}=9.185-0.313x-0.264x^2-0.321x^3,
\end{equation}
where $x$ is ${\log}R_{\rm 23}$.


\begin{figure}
\begin{center}
\includegraphics[width=10cm]{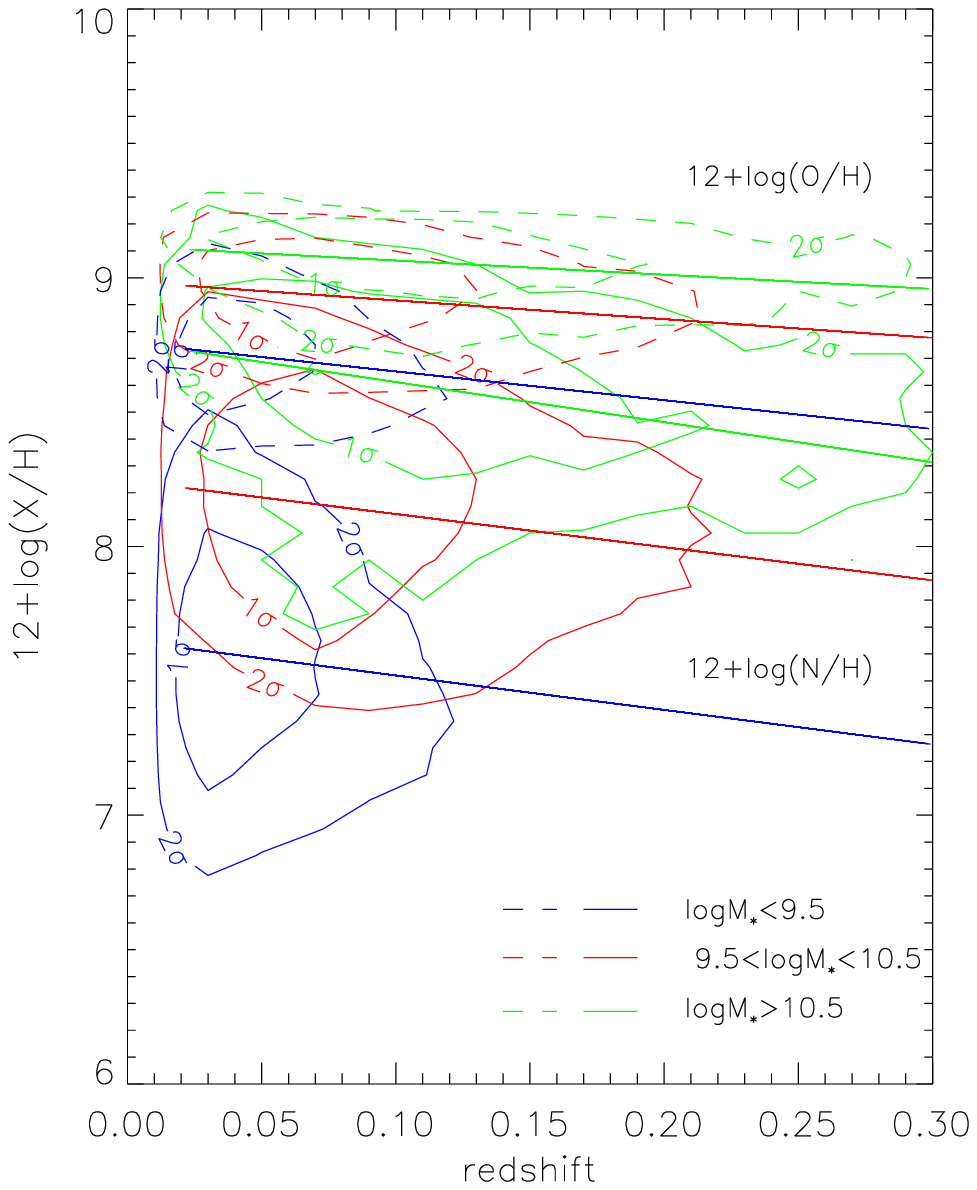}
\caption{Contours of oxygen and nitrogen abundances as a function of
redshift for those galaxies with $M_{*} < 10^{9.5} $, $10^{9.5}  -
10^{10.5} $, and $ > 10^{10.5} $. $1\sigma$ and $2\sigma$ correspond
to the $1\sigma$ and $2\sigma$ regions of the Gaussian distribution
for both redshift and 12+log(O/H) or 12+log(N/H). The dashed and solid curves are the contours of 12+log(O/H)
and 12+log(N/H), respectively. The blue, red, and green lines (both
solid and dashed) display those galaxies with $M_{*} < 10^{9.5} $,
$10^{9.5} - 10^{10.5}$, and $ > 10^{10.5}$, respectively. The
 straight lines (both solid and dashed)
are the best least-squares fits for these data.}
\end{center}
\end{figure}

The log(N/O) abundance ratios of these SFGs are estimated by using
the algorithm given by Thurston et al. (1996),
\begin{equation}
t_{\rm II}=6065+1600x+1878x^{2}+2803x^{3},
\end{equation}
where $t_{\rm II}$ is the \Nii temperature (in units of $10^{4}$ K), and $x$ is ${\log}R_{23}$.

Based on a five-level atom calculation, Pagel et al. (1992) given a
convenient formula:
\begin{equation}
{{\log} \frac{{\rm N}^{+}}{{\rm O}^{+}}}={{\log} \frac{{N}_{2}}{R_{2}}}+0.307-0.02 \rm log t_{\rm
II}-\frac{0.726}{t_{\rm II}}.
\end{equation}
Here, we take $N_{2}=1.33$ \Nii $\lambda 6584$ instead of the
standard $N_{2}$=\Nii $\lambda 6548$+\Nii $\lambda 6584$, because
the measurements of the \Nii $\lambda 6548$ line are less reliable
than those of the \Nii $\lambda 6584$ line (Thuan et al. 2010). To
derive N/O, we make the following assumption,
\begin{equation}
\frac{{\rm N}^{+}}{{\rm O}^{+}}=\frac{{\rm N}}{{\rm O}}.
\end{equation}
With regard to the accuracy of this assumption, some discussions
were given by Vila-Costas \& Edmunds (1993), and Thurston et al.
(1996) suggested that this assumption is fairly accurate.

The above calibration and procedure of obtaining the O abundance and
the N/O abundance ratio together for SFGs have been widely used in
literature, e.g., Liang et al. (2006), Mallery et al. (2007), and
Lara-L\'{o}pez et al. 2009. The total N abundance can then be
obtained from
\begin{equation}
\rm log\frac{\rm N}{\rm H}=\rm log\frac{{\rm N}}{{\rm O}}+\rm
log\frac{\rm O}{\rm H}.
\end{equation}

\section{RESULTS AND DISCUSSION}

In this section, we firstly present the galaxy downsizing effect,
i.e., enrichment ceases in higher-mass galaxies at earlier times and
shifts to lower-mass galaxies at later epochs (Pilyugin \& Thuan
2011); we show that both enrichment rates (i.e. slopes in Figs
$3-6$) and sSFRs (in Fig. 7) decrease between redshift 0.3 and
0.023, as well as that abundances at $z\sim0.023$ increase, but
sSFRs decrease with the galaxy stellar mass growth. Then we
demonstrate the O and N enrichments for $M_*>10^{11.0}$. Finally we
discuss the relation of the nitrogen productions between
intermediate-mass stars and massive stars using the model proposed
by Coziol et al. (1999) in SFGs.

\begin{figure}
\begin{center}
\includegraphics[width=10cm]{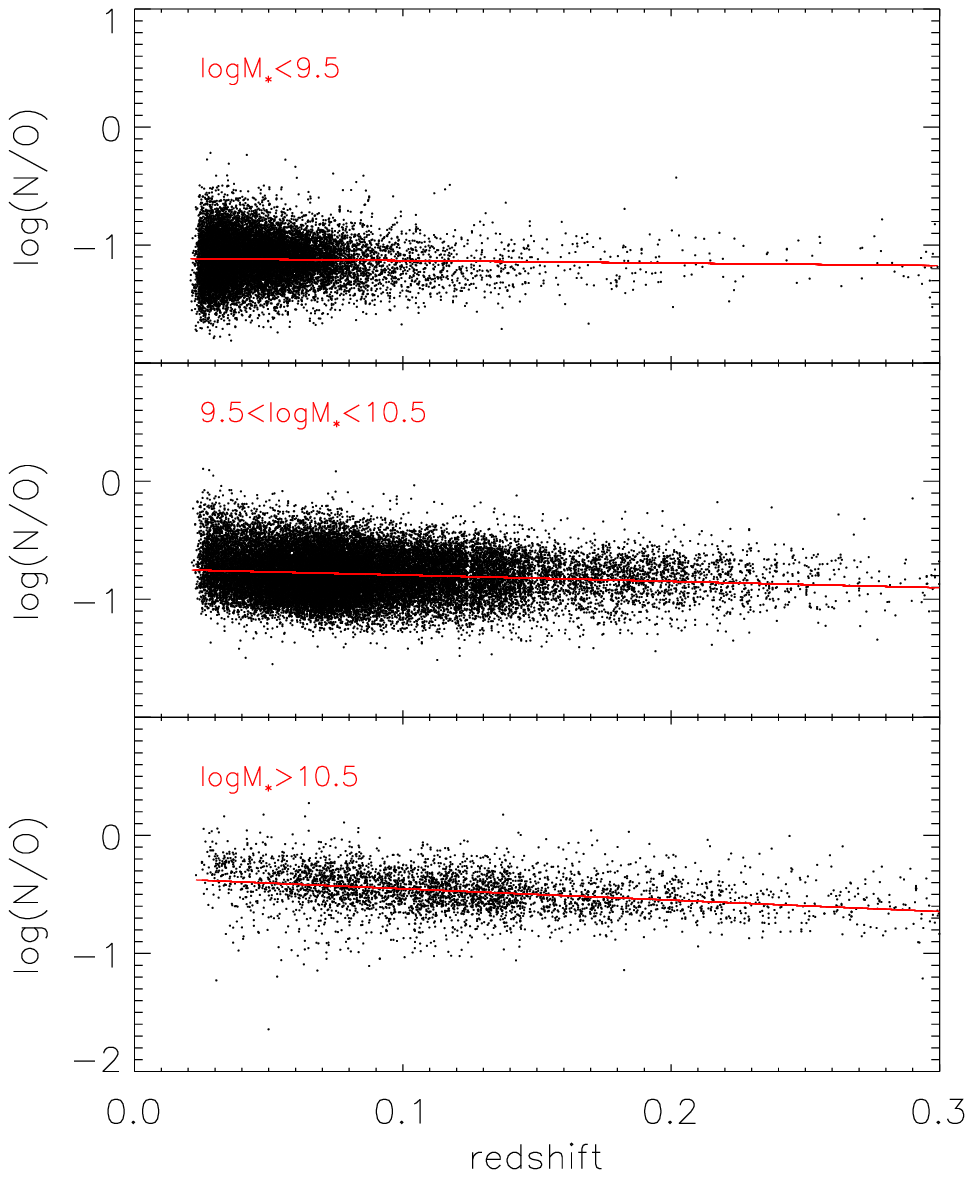}
\caption{Nitrogen-to-oxygen abundance ratios as a function of redshift for those galaxies with $M_{*} < 10^{9.5}
$, $10^{9.5}  - 10^{10.5} $, and $ > 10^{10.5} $. These abundance ratios are shown by black points. The red
solid lines are the best least-squares fits for these data.}
\end{center}
\end{figure}

\subsection{The Galaxy Downsizing Effect}

In Fig. 3, we show the redshift evolution of O and N abundances for
different galaxy mass ranges; their contours are presented in Fig.
4. In each panel, the red and black points represent O and N
abundances of SFGs at different redshifts, respectively, and the
best least-squares fits are shown by the solid lines. The upper
panel shows that the N and O abundance variations with redshift in
galaxies with $M_{*}< 10^{9.5}$. This indicates that these galaxies
have not reached a high astration level some 3 Gyr ago and are
experiencing intense star formation in their evolution processes.
This is consistent with the result of Pilyugin \& Thuan (2011).

\begin{table*}
\caption{Summary of nitrogen and oxygen abundances.}
\begin{small}
\begin{center}
\setlength{\tabcolsep}{6.5pt}
\renewcommand{\arraystretch}{0.8}
\begin{tabular}{lcccccccccccl}
\hline \hline Galaxy mass & sample size
&\multicolumn{3}{c}{log(N/O)}
& \multicolumn{3}{c}{12+log(N/H)}&\multicolumn{3}{c}{12+log(O/H)}\\
\cline{3-5} \cline{6-8} \cline{9-11} & & a & b & c
 & a &   b & c & a &   b & c
 \\
(1)& (2) & (3) & (4) &(5)&(6)&(7)&(8) & (9) & (10) & (11) \\
\hline
log$M_{*}\leqslant9.5$             & 17226     &-1.11  &-0.22 &0.06&7.65 & -1.30 &0.36&8.76  & -1.08 &0.30    \\
$9.5<\log M_{*}\leqslant10.5$      & 34100     &-0.74  &-0.57 &0.16&8.25 & -1.27 &0.35&8.99  & -0.71 &0.20    \\
log$M_{*}>10.5$                    & 3992      &-0.36  &-1.08 &0.30&8.76 & -1.64 &0.45&9.12  & -0.56 &0.16    \\
log$M_{*}>11.0$                    & 284       &-0.33  &-0.67 &0.19&8.77 & -1.01 &0.28&9.10  & -0.33 &0.09    \\
\hline \hline
\end{tabular}
\parbox{6.5in}
{\baselineskip 9pt \noindent \vglue 0.5cm {\sc Note}: Col.(1):
Galaxy mass. Col.(2): the sample size of SFGs. Cols.(3)-(5),
(6)-(8), and (9)-(11): For each sample of the log(N/O), 12+log(N/H),
and 12+log(O/H). ``a'' are the values of log(N/O), 12+log(N/H), and 12+log(O/H) at $z\sim0.023$, respectively. ``b'' are slopes (enrichment rates) of log(N/O), 12+log(N/H), and 12+log(O/H) between $z\sim 0.3$ and $z\sim0.023$, respectively.
 ``c" are $\Delta \rm (log(N/O))$, $\Delta \rm (12+log(N/H))$ , and $\Delta \rm (12+log(O/H))$ between $z\sim0.3$ and $z\sim0.023$ , respectively.}
\end{center}
\end{small}
\end{table*}

From Fig. 3 and Table 1, we can see that the slope for O abundance
increases from $-1.08$ to $-0.33$ when the stellar mass increases
from $M_{*}<10^{9.5}$ to $M_{*}<10^{11.0}$, as seen visually in Fig.
4. This indicates that the O enrichment (or star formation)
decreases with the galaxy stellar mass growth (see Table 1). This
demonstrates the galaxy downsizing effect, consistent with Pilyugin
\& Thuan (2011) who noted that O enrichment seems to decrease slowly
with increasing galaxy stellar mass. In addition, the O and N
abundances do not remain constant between different galaxy mass
ranges, and O abundance at $z\sim0.023$ increases from 8.76 to 9.12
with increasing galaxy stellar mass. This is inconsistent with the
result of Pilyugin \& Thuan (2011), who noted that O abundances are
always $\sim8.5$ for different stellar mass galaxies at
$z\sim0.023$. Moreover, the galaxies with the same stellar mass
range present significantly larger N enrichment than O enrichment
(see Table 1), which confirms the result of Thuan et al. (2010). It
is also noticeable that the difference between 12+log(O/H) and
12+log(N/H) decreases with the galaxy mass, indicating that the O
enrichment lags behind the N enrichment as galaxies grow, which will
be extensively discussed in section 4.4.

\begin{figure}
\begin{center}
\includegraphics[width=10cm]{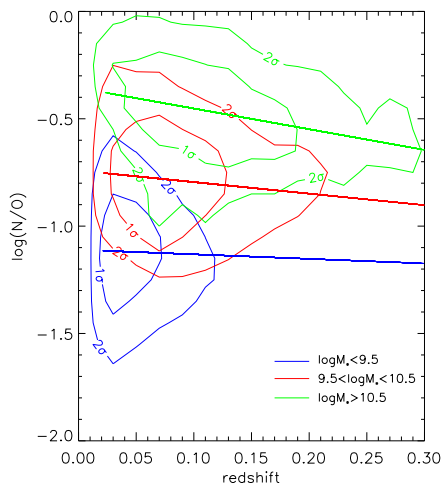}
\caption{Contours of nitrogen-to-oxygen abundance ratios as a function of redshift for those galaxies with
$M_{*} < 10^{9.5} $, $10^{9.5}  - 10^{10.5} $, and $ > 10^{10.5} $. $1\sigma$ and $2\sigma$ correspond to the
$1\sigma$ and $2\sigma$ regions of the Gaussian distribution for both redshift and log(N/O). The solid lines are
the best least-squares fits for these data.}
\end{center}
\end{figure}

In Fig. 5, we show the redshift evolution of N/O abundance ratios
for different galaxy mass ranges; their contours are presented in
Fig. 6. In the upper panel of Fig. 5, the sample size of galaxies
with $M_{*}< 10^{9.5}$ is 17226, and the N/O abundance ratio at
$z\sim0.023$ is $-1.11$. In the middle and lower panels of Fig. 5,
the sample sizes of galaxies with $M_{*}$ of $10^{9.5}-10^{10.5}$
and $>10^{10.5}$ are 34100 and 3992, respectively, and the N and O
abundances at $z\sim0.023$ are $-0.74$ and $-0.36$, respectively.
These indicates that the N/O abundance ratio increases with
increasing galaxy stellar mass. This is in good agreement with the
relation between N/O and the stellar mass of P\'{e}rez-Montero \&
Contini (2009).

Fig. 7 shows the redshift evolution of the specific star formation
rate (sSFR) for different galaxy mass ranges. In each panel, the
black points represent sSFRs of SFGs at different redshifts, and the
best least-squares fits are shown by the solid (red) lines. From
Fig. 7, we find that the slopes for sSFRs decrease from 6.39 to 4.39
and then to 3.58, and the intercepts for sSFRs at $z\sim0.023$
decrease from -9.70 to -10.00 and then to -10.11 with galaxy stellar
mass from $M_{*}< 10^{9.5}$ to $10^{9.5} - 10^{10.5}$ and then to
$>10^{10.5}$.  This is consistent with Fig. 3 of Pilyugin et al.
(2013). Clearly in this low redshift range the sSFRs of less massive
galaxies are on average always higher than that of more massive
galaxies, explaining the galaxy downsizing effect shown in Fig. 3 or
4. In other words, less massive galaxies are less massive because of
their less massive progenitors, despite of their higher sSFRs.

\subsection{The O and N Enrichments for Galaxies with $ M_{*}> 10^{11.0} $}

In this section, we firstly present the evidence of the O and N
enrichments for galaxies with $M_{*}>10^{11.0} $. Then we show the
redshift evolution of N/O abundance ratios for galaxies with
$M_{*}>10^{11.0} $. Finally, we discuss the O and N enrichments for
these most massive galaxies.

In Fig. 8, we show the O and N enrichments and the redshift
evolution of N/O abundance ratios for galaxies with
$M_{*}>10^{11.0}$. The sample size of galaxies with
$M_{*}>10^{11.0}$ is 284. The upper panel shows evidence of the N
and O enrichments. The slopes are $-0.33$ and $-1.01$ for O and N
abundances, respectively, and the N/O abundance ratios at
$z\sim0.023$ are $9.10$ and $8.77$, respectively. The lower panel
shows the redshift evolution of N/O abundance ratios. The slope is
$-0.67$ for N/O abundance ratios. In Fig. 13 of Thuan et al. (2010),
the O and N enrichments were shown for galaxies with $M_{*}$ of
$10^{10.0} -10^{10.3} $; however, the O and N enrichments were not
seen for galaxies with $M_{*}$ of $10^{11.2} -10^{11.5}$.

\begin{figure}
\begin{center}
\includegraphics[width=10cm]{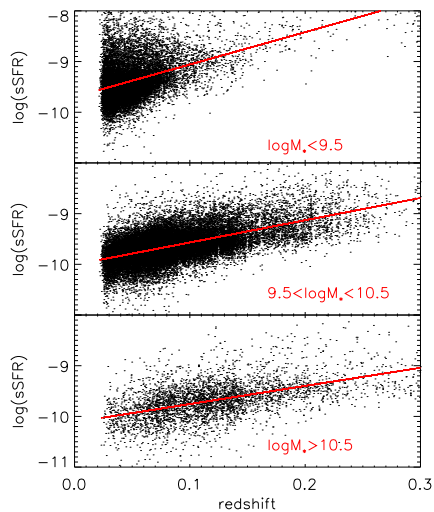}
\caption{Specific star formation rate (sSFR) as a function of redshift for galaxies with
$M_{*} < 10^{9.5} $, $10^{9.5}  - 10^{10.5} $, and $ > 10^{10.5} $.
The red solid lines are the best least-squares fits.}
\end{center}
\end{figure}

\begin{figure}
\begin{center}
\includegraphics[width=10cm]{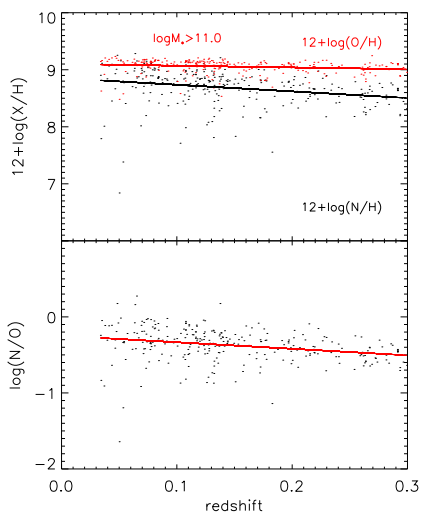}
\caption{O and N enrichments and N/O abundance ratio as a function of redshift for galaxies with $M_{*}
> 10^{11.0} $. Oxygen and nitrogen abundances are shown by red and black points, respectively, in
the upper panel. The abundance ratios are shown by black points in the lower panel. The red and black solid
lines are the best least-squares fits for these data.}
\end{center}
\end{figure}

Employing data of the VIMOS VLT Deep Survey, Lamareille et al.
(2009) obtained the mass-metallicity relation of SFGs and have found
that the galaxies of $10^{10.2} $ show a larger O enrichment than
those of $10^{9.4} $. In addition, Lara-L\'{o}pez et al. (2009) have
studied the O abundance of relatively massive (log$(M_{*})\geqslant
10.5$) SFGs from SDSS/DR5 at different redshift intervals from 0.4
to 0.04. They found an oxygen enrichment $\Delta({\log}({\rm O/H}))
\sim 0.1$ from redshift 0.4 to 0. From Fig. 9 and Table 1, we can
see clearly $\Delta({\log}({\rm O/H})) \sim 0.10$ and
$\Delta({\log}({\rm N/H})) \sim 0.28$ from redshift 0.023 to 0.30.
Comparing with the above results, we may safely conclude that the O
and N are enriched for galaxies with $M_{*}>10^{11.0}$.

\subsection{The N Production in Star Forming Galaxies}

\begin{figure}[t]
\begin{center}
\includegraphics[width=10cm]{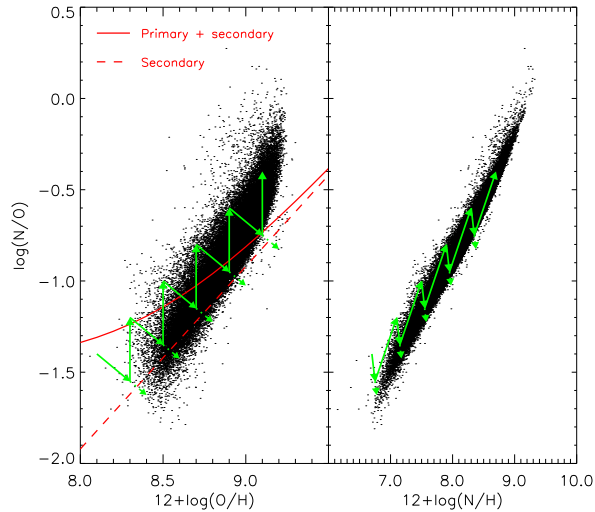}
\caption{Schematic demonstration of the nitrogen productions of
intermediate-mass and massive stars in a sequence of bursts. During
a cycle, the massive star evolution not only produces the oxygen
enrichment, but also produces the nitrogen enrichment; the
intermediate-mass star evolution produces only the nitrogen
enrichment. In all cycles, the enrichment of the intermediate-mass
stars should be larger than that of the massive stars.}
\end{center}
\end{figure}

In this section, we first introduce a schematic demonstration of
productions of primary and secondary nitrogen in a sequence of
bursts (Coziol et al. 1999). Then we show that the nitrogen
production of intermediate-mass stars is larger than that of massive
stars in SFGs.

Based on the evolutionary schematic model of N/O ratios in Garnett
(1990), Coziol et al. (1999) proposed a scheme that the production
of nitrogen is the consequence of a sequence of bursts in SFGs. This
means that the N abundance and the metallicity increase in SFGs,
because these galaxies experience a couple of star formation
processes in several cycles. During a cycle, the $12+{\log}({\rm
O/H})$ increases due to the evolution of massive stars, while the
N/O ratio decreases (Garnett 1990; Olofsson 1995; Coziol et al.
1999). After $\sim 0.4$ Gyrs of massive star active onset, the
intermediate-mass stars contribute most nitrogen, and the N/O ratio
increases sharply, while the $12+{\log}({\rm O/H})$ does not
increase.

\begin{figure}
\begin{center}
\includegraphics[width=10cm]{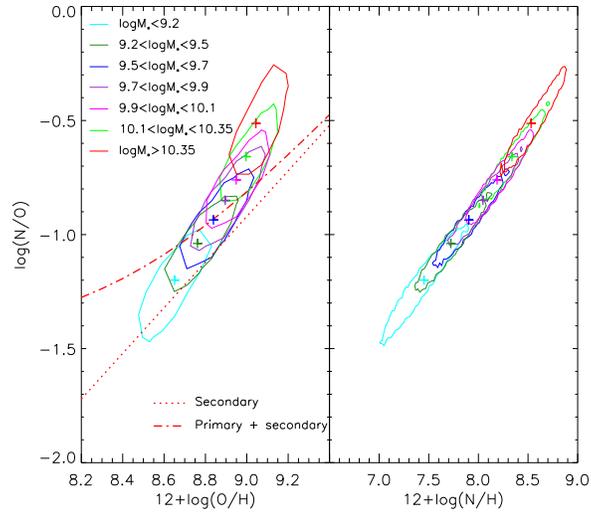}
\caption{Contours of N/O abundance ratios as functions of 12+log(O/H) and 12+log(N/H) for those galaxies with
different galaxy mass ranges. $1\sigma$ corresponds to the $1\sigma$ region of the Gaussian distribution for
both log(N/O), 12+log(O/H), and 12+log(N/H). Seven `+' signs mark the central values of the distributions for
these parameters among seven different galaxy mass ranges.}
\end{center}
\end{figure}

Utilizing the SDSS DR7 data, Torres-Papaqui et al.
(2012) found that the intensity of the bursts seems to result in the
chemical differences between the nitrogen-poor and nitrogen-rich
SFGs, which is well explained by the sequence of bursts model
(Coziol et al. 1999). In addition, the SDSS DR8 data
are selected as the SFG sample, and Wu \& Zhang (in preparation) found that
the metallicity increase on the BPT diagram can be explained by
several star formation processes. These suggest that a sequence of
bursts model may help to understand the chemical evolution of SFGs.

In Fig. 9, we show the relations of ${\log}({\rm N/O})$ versus
$12+{\log} ({\rm O/H})$ and $12+{\log} ({\rm N/H})$, respectively.
Generally, the N/O ratio will be constant for primary
nucleosynthesis, while this ratio will be a linear correlation for
secondary nucleosynthesis. The combination of both primary and
secondary nucleosyntheses gives rise to a nonlinear relation
(Mallery et al. 2007). During the first cycle, the evolution of
massive stars contribute not only the $12+{\log}({\rm O/H})$
increase but also the N/O ratio decrease in the left panel; in the
meantime, this process also produces some nitrogen (the right panel
in Fig. 9). With the evolution of these galaxies, the massive stars
start dying off, and the intermediate-mass stars open large scale
nitrogen production (the right panel in Fig. 9), but the oxygen
production ceases (the left panel in Fig.9). In the right panel, it
can be seen that the nitrogen production of the intermediate-mass
stars is larger than that of the massive stars. Therefore, the
nitrogen production is dominated by the intermediate-mass stars in
SFGs during the first cycle.

\begin{figure}
\begin{center}
\includegraphics[width=10cm]{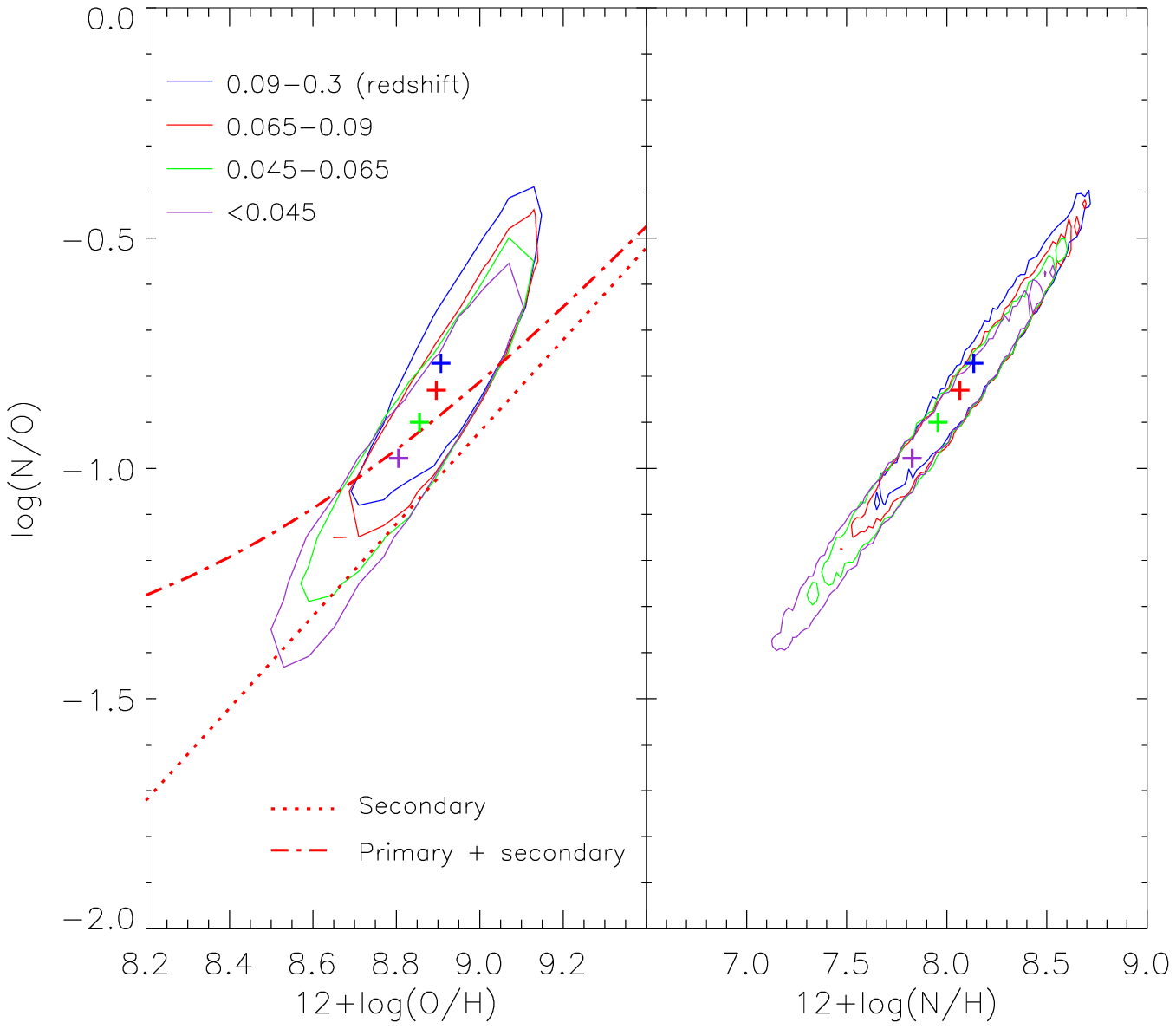}
\caption{Contours of N/O abundance ratios as functions of
12+log(O/H) and 12+log(N/H) for those galaxies with different
redshift ranges. $1\sigma$ corresponds to the $1\sigma$ region of
the Gaussian distribution for both log(N/O), 12+log(O/H), and
12+log(N/H). Four pluses are the central values of the distributions
for these parameters among four different redshift ranges.}
\end{center}
\end{figure}

The second cycle begins with an increase in the $12+{\log}({\rm
O/H})$ and a decreases in the N/O ratio, and this cycle will follow
the first cycle regardless of that the star formation process has a
lower or higher amplitude (intensity) than that of the first cycle.
Actually, some models of successive bursts assume that the bursts
will have decreasing intensities (Gerola, Seiden, \& Schulman 1980;
Krugel \& Tutukov 1993; Marconi, Matteuci, \& Tosi 1994; Koeppen,
Theis, \& Hensler 1995; Coziol et al. 1999). However, our result
that the nitrogen production is dominated by the intermediate-mass
stars in SFGs will not be changed, because the nitrogen production
processes of intermediate-mass and massive stars are the same as
those of the first cycle.

Regarding relative high-metallicities ($12 + {{\log}({\rm O/H})}$
$\gtrsim 8.3$) objects, the N/O ratio increases significantly with
the O abundance growth. The trend seems to originate from the
nitrogen production with metallicity-dependence in both massive and
intermediate-mass stars (e.g., Vila-Costas \& Edmunds 1993; Pilyugin
et al. 2003). Therefore, the nitrogen is generally a secondary
element in the metallicity range (Vila-Costas \& Edmunds 1993;
Henry, Edmunds, \& K\"{o}ppen 2000; van Zee \& Haynes 2006;
L\'{o}pez-S\'{a}nchez \& Esteban 2010). Since primary synthesis is
commonly assumed (but not definitely) to confine in
intermediate-mass stars, while secondary synthesis could occur in
stars of all masses (Renini \& Voli 1981; Vila-Costas \& Edmunds
1993), our result that the nitrogen production is dominated by the
intermediate-mass stars is reasonable. During the main sequence
(MS), intermediate mass stars burn hydrogen through the CNO cycle.
The conversion of almost all central $\rm{C}^{12}$ into
$\rm{N}^{14}$ is the consequence of the CNO cycle (Smiljanic et al.
2006). In addition, intermediate-mass stars seem most likely to
contribute, during the asymptotic giant branch, the CNO-processed
material (Cannon et al. 1998).

In term of initial mass function (IMF), equations
 1 and 2 in Kroupa (2001) show that
a mean stellar mass $\langle m\rangle$=0.36$M_{\sun}$ for stars
with 0.01 $\leq$ m $\leq$ 50 $M_{\sun}$, and 5.7 \%
`intermediate-mass (IM) stars' (1.0-8.0 $M_{\sun}$) contribute
34 \% mass, and 0.37 \% `O' stars ($>$8.0 $M_{\sun}$)
contribute 17 \% mass (Kroupa 2001).
Therefore intermediate-mass stars can contribute twice as much
mass as massive star in SFGs. This is consistent  intermediate-mass stars
dominate the nitrogen production in SFGs.

\subsection{Enhanced N/O ratio in massive galaxies}

However, most of the galaxies in our sample are located
significantly above the secondary synthesis line, which has been
proposed to dominate the total synthesis of N/O. It is evident that
the last arrows that marks the nitrogen production by the
intermediate-mass stars in both panels point to the direction of the
observed N/O enhancement. This means that no O abundance enhancement
is needed for the following star-forming processes. This is
inconsistent with Coziol et al. (1999), who suggested that the sum
of the vectors of alternating oxygen (by massive stars) and nitrogen
(by intermediate-mass stars) converges to the secondary synthesis
line. Due to the existence of the outflow/winds, the separation in
Fig. 9 may originate from the galactic wind driven by the starburst
(Strickland, Ponman, \& Stevens 1997; Moran, Lehnert, \& Helfand
1999). The outflow will preferentially deplete
 the oxygen, and nitrogen will remain at its ISM value at the time of
  the starburst/wind (Lehnert \& Heckman 1996; Heckman 2003;
  Torres-Papaqui et al. 2012). Therefore, the outflow effect may result
   in the separation of the metallicities of the galaxies from the
  secondary + primary chemical evolution model of
   Vila-Costas \& Edmunds (1993).

In order to understand the above discrepancy, in Figs 10 and 11 we
show their contours and central values (with `+' sign) for different
galaxy mass and redshift ranges. Each contour includes 68.3\% of the
total galaxies for this subsample; all subsamples have almost an
equal size in each Figure. Fig. 10 shows clearly that larger mass
galaxies show more significant deviations of N/O abundance ratio
above the secondary synthesis line. This is in good agreement with
the relation between N/O abundance ratio and O abundance with the
stellar mass of Amor\'{\i}n, P\'{e}rez-Montero, \& V\'{\i}lchez
(2010). Fig. 11 shows that higher redshift galaxies have
statistically larger N/O abundance ratios than lower redshift ones.
Using a good indicator of the stellar age, Kauffmann et al. (2003b)
displayed a correlation between the $D_{n}(4000)$ index and stellar
mass; this relation indicates that massive galaxies seem to be older
(Tremonti et al. 2004). This means that higher redshift and more
massive galaxies have started their nitrogen enrichment earlier;
this is another manifestation of the downsizing effect. Since the N
abundance enrichment process is not accompanied with any significant
oxygen enrichment, we conclude that the outflows of massive stars,
which
 deplete oxygen efficiently, are more significant in massive galaxies.

In Table 2, we list the sample size, mean and median values of
$M_{*}$, $\log\frac{\rm N}{\rm O}$, $12+\log\frac{\rm N}{\rm H}$ and
$12+\log\frac{\rm O}{\rm H}$ of these galaxies in each subsample
shown in Figs 10 and 11; in each case, the mean and median values
are approximately the same. In Fig 12, we show the relations
between the median values of $M_{*}$ and $\log\frac{\rm N}{\rm O}$,
$12+\log\frac{\rm N}{\rm H}$ and $12+\log\frac{\rm O}{\rm H}$'
respectively; a linear fit to each relation is also shown as the
dashed line. Clearly a good linear relation exists between $\log
M_{*}$ and $\log\frac{\rm N}{\rm O}$, i.e.,
\begin{equation}
\log\frac{\rm N}{\rm O}=-5.09+0.43\log M_{*},
\end{equation}
suggesting that N/O abundance ratio is a statistically excellent
indicator of the mean/median stellar mass of a sample of SFG. Since
$M_*$ is derived from the total stellar lights of a galaxy, a
cosmological model (luminosity distance) must be assumed when
calculating $M_*$ from the multi-band photometric data. The ability
of predicting reliably $M_*$ with Equation (7) from the observed N/O
abundance ratio, which is cosmological model independent, suggests
that the relation in Equation (7) can be used a standard candle to
study cosmology. The parameters (slope and intercept) in Equation
(7) are now determined with a given cosmological model in the
catalog, thus cannot be used directly to study alternative
cosmological models. However, we can use the observed SNe Ia, which
are excellent standard candles, to calibrate the relation in
Equation (7), in a model-independent way similar to that the method
of calibrating several luminosity relations of gamma-ray bursts
(Liang et al. 2008, 2010). Since the relationship in equation (7) is
 derived using only the median values of N/O abundance ratio, its
  statistical robustness should be carefully studied in the future
   with, e.g. the method of Kelly (2007). Finally, this relationship
    needs to be tested in a broader redshift range before it is applied
     for cosmological studies. However further more detailed discussion
on this subject is beyond the scope of this present work and will be
presented elsewhere.

With the observational data of 55,318 SFGs selected
from the catalog of MPA-JHU emission-line measurements for the SDSS
DR8, we find
 evidence of the galaxy downsizing effect,
the O and N enrichments for galaxies with stellar masses larger than
$10^{11.0}$, and the nitrogen production dominated by the
intermediate-mass stars in SFGs. We summarize our main results
below.

\begin{table*}
\caption{Summary of SFGs.}
\begin{small}
\begin{center}
\setlength{\tabcolsep}{1.5pt}
\renewcommand{\arraystretch}{1.2}
\begin{tabular}{lccccccccccccl}
\hline \hline Parameters & sample size &\multicolumn{2}{c}{$\log
M_{*}$} &\multicolumn{2}{c}{log(N/O)}
& \multicolumn{2}{c}{12+log(N/H)}&\multicolumn{2}{c}{12+log(O/H)}\\
\cline{3-4} \cline{5-6} \cline{7-8} \cline{9-10} & & mean & median
 & mean &  median & mean &  median  & mean &  median
 \\
(1)& (2) & (3) & (4) &(5)&(6)&(7)&(8) & (9) & (10)  \\
\hline
$\log M_{*}\leqslant9.2$       & 8734 &8.85  &8.95  &-1.20&-1.21  &7.45 &7.43  &8.65 & 8.64  \\
$9.2<\log M_{*}\leqslant9.5$   & 8492 &9.36  &9.37  &-0.97&-1.04  &7.72 &7.72  &8.81 & 8.76  \\
$9.5<\log M_{*}\leqslant9.7$   & 7974 &9.60  &9.61  &-0.94&-0.94  &7.90 &7.90  &8.84 & 8.85  \\
$9.7<\log M_{*}\leqslant9.9$   & 8487 &9.80  &9.80  &-0.85&-0.85  &8.05 &8.05  &8.90 & 8.91  \\
$9.9<\log M_{*}\leqslant10.1$  & 7551 &10.00 &9.99  &-0.76&-0.77  &8.19 &8.19  &8.95 & 8.96  \\
$10.1<\log M_{*}\leq10.35$     & 7286 &10.22 &10.21 &-0.66&-0.67  &8.34 &8.34  &9.00 & 9.01  \\
$\log M_{*}>10.35$             & 6794 &10.58 &10.53 &-0.53&-0.53  &8.51 &8.53  &9.04 & 9.06  \\
\hline \hline Parameters & sample size &\multicolumn{2}{c}{z}
&\multicolumn{2}{c}{log(N/O)}
& \multicolumn{2}{c}{12+log(N/H)}&\multicolumn{2}{c}{12+log(O/H)}\\
\cline{3-4} \cline{5-6} \cline{7-8} \cline{9-10} & & mean & median
 & mean &  median & mean &  median  & mean &  median\\
 \hline
$z\leqslant0.045$              &13732 &0.03&0.03&-0.98&-1.01  &7.83 &7.79  &8.80 &8.79  \\
$0.045<z\leqslant0.065$        &13101 &0.06&0.05&-0.90&-0.91  &7.95 &7.94  &8.85 &8.86  \\
$0.065<z\leqslant0.09$         &13462 &0.08&0.08&-0.83&-0.84  &8.06 &8.08  &8.89 &8.91  \\
$0.09<z<0.3$                   &15023 &0.13&0.12&-0.78&-0.78  &8.12 &8.16  &8.90 &8.93  \\

\hline \hline
\end{tabular}
\parbox{6.5in}
{\baselineskip 9pt \noindent \vglue 0.5cm {\sc Note}: Col.(1):
$M_{*}$ or redshift (z) ranges. Col.(2): the subsample size of SFGs.
Cols.(3)-(4), (5)-(6), (7)-(8), and (9)-(10): {``mean''} and {``median''}
are the mean and median values of $M_{*}$ or z, log(N/O),
12+log(N/H), and 12+log(O/H), respectively, in each subsample.}
\end{center}
\end{small}
\end{table*}

\begin{figure}
\begin{center}
\includegraphics[width=10cm]{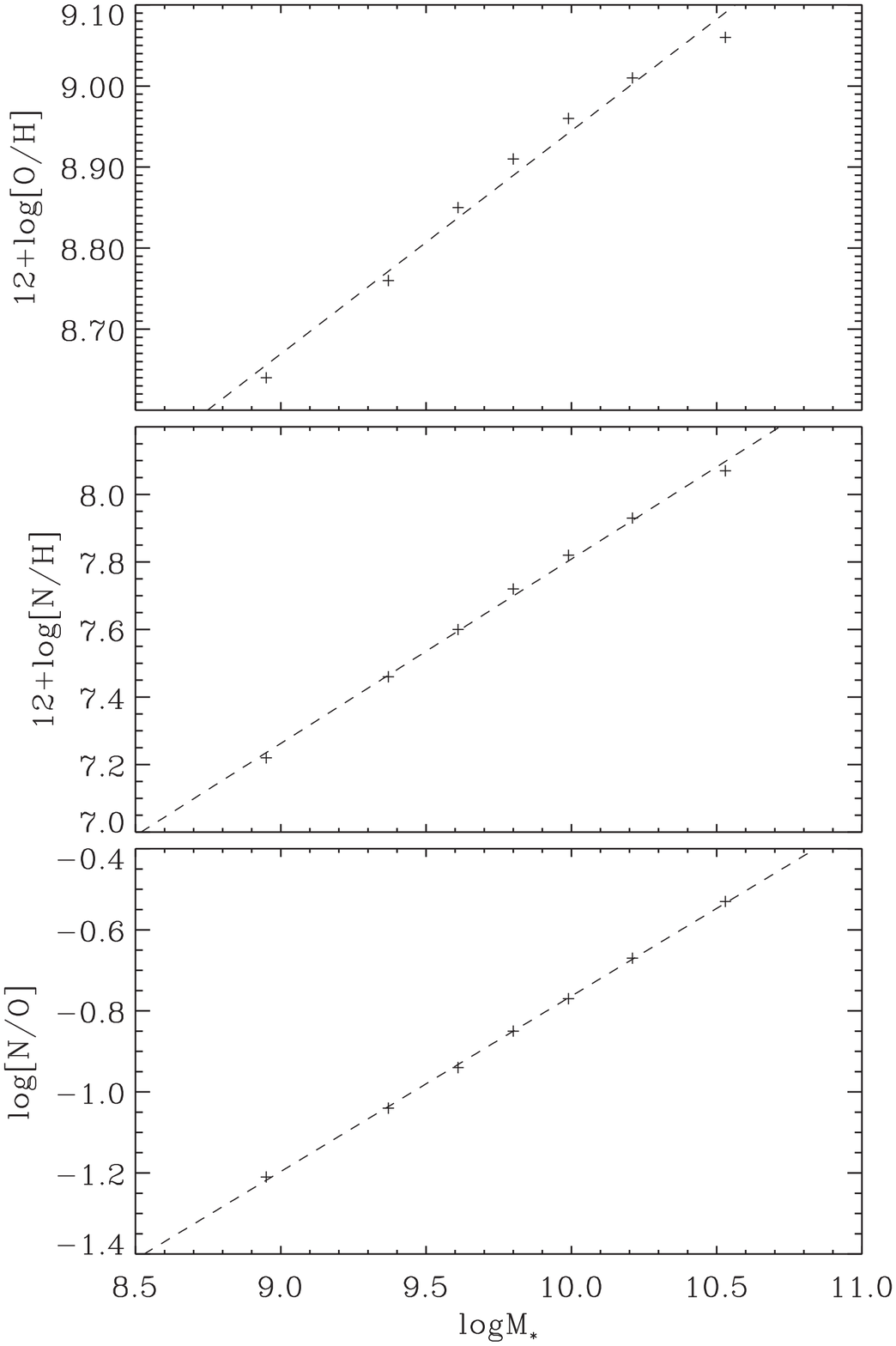}
\caption{Relations between the median values of $M_{*}$ and
$\log\frac{\rm N}{\rm O}$ (bottom panel)£¬ $12+\log\frac{\rm
N}{\rm H}$ (middle panel) $12+\log\frac{\rm
O}{\rm H}$ (top panel) for each subsample shown in Fig. 9.}
\end{center}
\end{figure}


(1) We show the redshift evolution of O and N abundances for
different galaxy mass ranges, and it presents the galaxy downsizing
 effect, consistent with Pilyugin \& Thuan (2011).  We also show that
  in this low redshift range the sSFRs of less massive galaxies are on
   average always higher than that of more massive galaxies, consistent
    with Pilyugin et al. (2013). This explains the galaxy downsizing
     effect, i.e., less massive galaxies are less massive because of
      their less massive progenitors, despite of their higher sSFRs.

(2) The O and N abundances do not remain constant at different
galaxy mass ranges, and the enrichment capability (SFRs) decreases
with the galaxy stellar mass growth. The O abundance at
$z\sim0.023$'' increase
 from 8.76 to 8.99 and then to 9.12 with
increasing galaxy stellar mass, which is inconsistent with the
result of Pilyugin \& Thuan (2011). Moreover, the galaxies with the
same stellar mass range present significantly larger N enrichment
than O enrichment (Table 1), which confirms the result of Thuan et
al. (2010).

(3) We show the redshift evolution of N/O abundance ratios for
different galaxy mass ranges. We find N/O abundance ratios at
$z\sim0.023$ increase with the galaxy stellar mass growth, and the
slopes decrease with the galaxy stellar mass growth. This is in good
agreement with P\'{e}rez-Montero \& Contini (2009) and Amor\'{\i}n,
P\'{e}rez-Montero, \& V\'{\i}lchez (2010).

(4) For the first time we find evidence of the O and N enrichments
for galaxies with $M_{*}>10^{11.0} $. In contrast to previous
conclusion that the most massive galaxies do not show an appreciable
enrichment in oxygen, we find  $\Delta({\log}({\rm O/H})) \sim 0.10$
and $\Delta({\log}({\rm N/H})) \sim 0.28$ from redshift 0.023 to
0.30 for these very massive galaxies.

(5) We conclude that the nitrogen production is dominated by the
intermediate-mass stars, which dominate the secondary synthesis in
SFGs.

(6) We find that the N/O abundance ratios of SFGs with
$M_{*}>10^{10.35}$ are located significantly above the secondary
synthesis line. This suggests that outflows of massive stars, which
deplete oxygen efficiently, are more important in massive galaxies.

(7) We find an excellent linear relation between $M_{*}$ and
log(N/O), indicating that the N/O abundance ratio is good indicator
of the stellar mass in a SFG, which may be used as a standard candle
for studying cosmology after proper calibration with some other
cosmology independent standard candles, such as SNe Ia. However,
further careful studies are needed before it is applied to
cosmological studies.

\acknowledgments

YZW thanks Yanchun Liang for valuable discussions. The anonymous referee is thanked for many constructive comments and suggestions, which allowed us to improve the paper significantly. SNZ acknowledges partial funding support by 973 Program of
China under grant 2009CB824800, by the National Natural Science Foundation of China under grant Nos. 11133002
and 10725313, and by the Qianren start-up grant 292012312D1117210.

\end{document}